\documentclass[pre,preprint, groupedaddress, showpacs]{revtex4}

\usepackage{graphicx}
\begin{document}


\title{Cycloid motions of grains in unmagnetized dust plasma}


\author{Ya-feng He}
\email[Email:]{heyf@hbu.edu.cn}
\author{Wei-hua Gong}
\author{Yong-liang Zhang}
\author{Fu-cheng Liu}


\affiliation{Hebei Key Laboratory of Optic-electronic Information Materials, College of Physics Science and Technology, Hebei University, Baoding 071002, China.}


\date{\today}
\begin{abstract}
\indent Hypocycloid and epicycloid motions of irregular grain (pine pollen) are observed for the first time in unmagnetized dust plasma in 2D horizontal plane. Hypocycloid motions occur both inside and outside the glass ring which confines the grain. Epicycloid motion only appears outside the glass ring. Cuspate cycloid motions, circle motion, and stationary grain are also observed. All these motions are related with both the initial conditions of dropped grain and the discharge parameters. The Magnus force originated from the spin of the irregular grain is confirmed by comparison experiments with regular microspheres, and it plays important role on these (cuspate) cycloid motions. The observed complex motions are explained in term of force analysis and numerical simulations. Periodical change of the cyclotron radius as the grain travelling results in the (cuspate) cycloid motions. Our results show that the (cuspate) cycloid motions are distinctive features of irregular grain immersed in plasma.
\end{abstract}



\maketitle
\section {Introduction}
\indent Dust plasma is characterized by charged grains immersed in a plasma of electrons and ions \cite{Morfill,Fortov1,Fortov2,Shukla1}. The grains could acquire charges of the order of $10^{3}$ to $10^{4}$ elementary charges and often become strongly coupled, which gives rise to novel phenomena such as the crystallization, phase transitions, and waves \cite{Chu, Zhdanov, Shukla2, Kalman, Arp, Bonitz1}. The grains suspended in sheath suffer gravitation force, electrostatic force, drag force, and sometime external magnetic/laser field etc., which leads to complex motions of grains like gyromotion, cluster rotation and shear flow \cite{Ott, Feng, Nosenko2, Davoudabadi, Carstensen, Amatucci, Ishihara, Karasev, Saitou, Hou, Wang, Kaw, Bonitz2, Baruah, Konopka}. It is obvious that a charged grain could perform gyromotion if it is magnetized in external magnetic field. The measurements of gyroradius and cyclotron frequency provide an effective method for the noninvasive determination of the charge on a grain \cite{Amatucci}.

\indent Most of the studies on the dust plasma both in experiment and in theory deal with grain of a spherical form, such as the glass, melamine formaldehyde, and alumina mocrosphere \cite{Chu, Zhdanov,Amatucci}. In some cases, especially in the numerical simulation of molecular dynamics, for simplicity, the grain is considered as a charged particle neglecting the grain shape. However, the grains appeared in planetary rings, fusion reactors, materials processing, etc., generally have a nonspherical shape. The dynamics of the nonspherical grain in dust plasma is less known at present.

\indent A grain immersed in a plasma suffers the momenta transferred from plasma flux and surrounding gas \cite{Fortov1,Morfill,Fortov2,Shukla1}. This could result in the spin of the grain and has been paid much attentions in recent years \cite{Tsytovich, Karasev2, Karasev3, Hutchinson, Krasheninnikov, Paeva, Dzlieva1, Dzlieva2}. For an ideal spherical grain, the net effect for the momenta transfer can be zero. For a grain with irregular shape the net transfer of the angular momenta can be estimated as $\eta_{ass}\pi a^{2}n_{i,\infty}m_{i}av_{s}v_{\infty}$ \cite{Tsytovich}, where $\eta_{ass}$ is the coefficient taking into account the degree of irregularity of the grain shape. The angular frequency of such spin can reach a rather large values ($\sim 10^{4}-10^{9} s^{-1}$) \cite{Karasev2, Karasev3, Harwitt, Mahmoodi}. The spin of hollow glass microsphere with defect has been detected with angular velocity up to 12000 rad/s in a stratified glow discharge \cite{Karasev2, Karasev3}. The spin of a charged grain could give rise to a magnetic moment. Interacting with external magnetic field, the spinning grain is subjected to a magnetic force which could result in the complex motions of such grain \cite{Tsytovich, Krasheninnikov, Dzlieva2, Schwabe}. However, in the absent of external magnetic field, a spinning grain could also exhibit complex motions which have not been well studied until now. Here, we report a study on the cycloid motions of irregular grain (pine pollen) in unmagnetized dusty plasmas. Hypocycloid and epicycloid motions are observed, besides the stationary grain and the circle motion of grain. The Magnus force originated from the spin of the grain is first considered, and it plays important role on the cycloid motions of grains. These complex motions of grains are explained based on force analysis and numerical simulations in 2D horizontal plane. Additional comparison experiments with regular microspheres are performed in order to confirmed that the cycloid motions are distinctive features of a pine pollen immersed in plasma.

\section{EXPERIMENTAL SETUP}

\indent Figure 1 shows the sketch of the experimental apparatus. The experiments are carried out in a vacuum chamber with a background pressure within $p$$=$$10-70$ Pa. Argon gas at flow of $10$ sccm is used for the discharge. In order to avoid disturbances in grain motion during an experiment the position of the gas inlet is designed properly. Plasma is produced by coupling capacitively the electrode to a radio-frequency generator ($13.56$ MHz). The forward power varies from $5$ to $50$ W. The separation between the lower stainless steel electrode and the upper ITO coated electrode is $70$ mm.

\begin{figure}[htbp]
  \begin{center}\includegraphics[width=6cm,height=7cm]{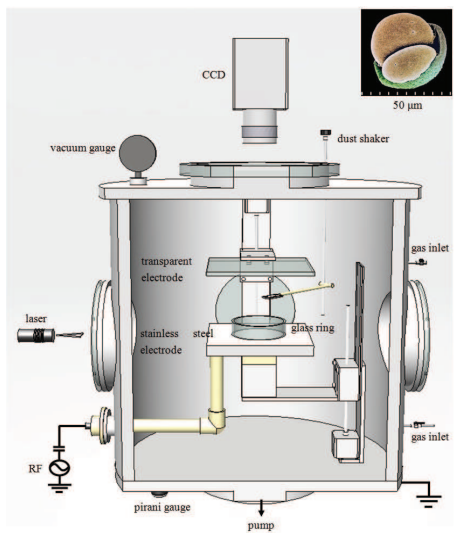}
  \caption{(Color online) Schematic diagram of the experimental setup. The inset shows the image of pine pollen we used in experiments.}
  \end{center}
\end{figure}

\indent Pine pollen having nonspherical shape as shown by the inset in Fig. 1 are injected into the plasma through a shaker and serve as dust grains. Their average radii are about $20$$\pm$$5$ $\mu$m. One or few grains are injected into the discharge during each experiment in order to avoid the formation of grains ring and plasma condensation. Regular glass and resin microspheres are also used in comparison experiments. Grain is horizontally illuminated by a $50$ mW, $532$ nm laser sheet with a thickness of $0.5$ mm and imaged by a CCD camera from the upper transparent electrode. The spatial and temporal resolutions are $10$ $\mu$m/pixel and $50$ frames/s, respectively. The trajectory of grain can be obtained by using long-time exposure. Characteristic parameters of the grain movement are measured by using the image processing with MATLAB.

\indent A glass ring is positioned on the lower stainless steel electrode. The diameter and the height of the glass ring are $37$ mm and $10$ mm, respectively. After being injected into the plasma, charged grain levitates nearby the glass wall. Due to the circular symmetry of the glass ring the grain travels around the glass ring or stays near its equilibrium position.

\section{EXPERIMENTAL RESULTS}
\subsection{Regular movements of grain }
\indent We firstly show the regular movement of a single grain around the glass ring in 2D horizontal plane as shown in Fig. 2. Figure 2(a) shows an epicycloid motion with inward petals. This meandering motion can be regarded as a compound of two circular motions, where the primary circle (radius $r_{2}$) orbits the secondary circle (radius $r_{1}$) in one direction with angular velocity $\omega_{1}$ and spins about its center in the same direction with angular velocity $\omega_{2}$ as shown by the diagram of Fig. 3(a). Figure 2(c) shows a hypocycloid motion with outward petals with rotations in the opposite sense [as illustrated by the diagram of Fig. 3(b)]. The epicycloid and hypocycloid motions occur in two cases, respectively. The epicycloid trajectory is observed only outside of the glass ring, and the value of angular velocity $\omega_{1}\omega_{2}$$>$$0$. The hypocycloid trajectory is found both inside and outside the glass ring, and the value of angular velocity $\omega_{1}\omega_{2}$$<$$0$. For the two kinds of cycloid motions they both satisfy the following relation:
\begin{eqnarray}
  |r_{1}\omega_{1}|<|r_{2}\omega_{2}|,
\end{eqnarray}
i.e., the linear speed of the secondary circle is less than that of the primary circle. This is a key to the formations of the petals on both epicycloid and hypocycloid trajectories.

\begin{figure}[htbp]
  \begin{center}\includegraphics[width=8cm,height=6cm]{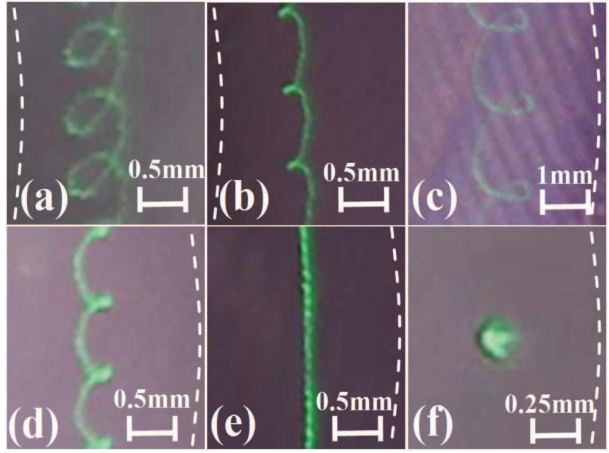}
  \caption{(Color online) Trajectories of travelling (a)-(e) and stationary (f) grains. Dashed line in each image indicates schematically the glass ring. (a) and (b) are located outside the glass ring. (c)-(f) are located inside the glass ring. (a): epicycloid motion; (b): cuspate epicycloid motion; (c): hypocycloid motion; (d): cuspate hypocycloid motion; (e): circular motion; (f): stationary grain. Exposure times of (a)-(f) are: 0.39, 0.26, 0.53, 0.16, 0.32, 1 s, respectively.}
  \end{center}
\end{figure}

\indent Cuspate epicycloid and hypocycloid motions are observed as shown in Fig. 2(b) and 2(d), respectively. In these cases, the radius $r_{1,2}$ and the angular velocity $\omega_{1,2}$ follow the condition: \begin{eqnarray}
  |r_{1}\omega_{1}|\geq|r_{2}\omega_{2}|.
\end{eqnarray}
 This means that the linear speed of the secondary circle is equal or greater than that of the primary circle, which leads to the appearance of the cusps on the cuspate cycloid trajectories. The cuspate epicycloid motion is observed only outside the glass ring, and the value of angular velocity $\omega_{1}\omega_{2}$$>$$0$. The cuspate hypocycloid motion is observed both inside and outside the glass ring, and the value of angular velocity $\omega_{1}\omega_{2}$$<$$0$.

\begin{figure}[htbp]
  \begin{center}\includegraphics[width=8cm,height=3.9cm]{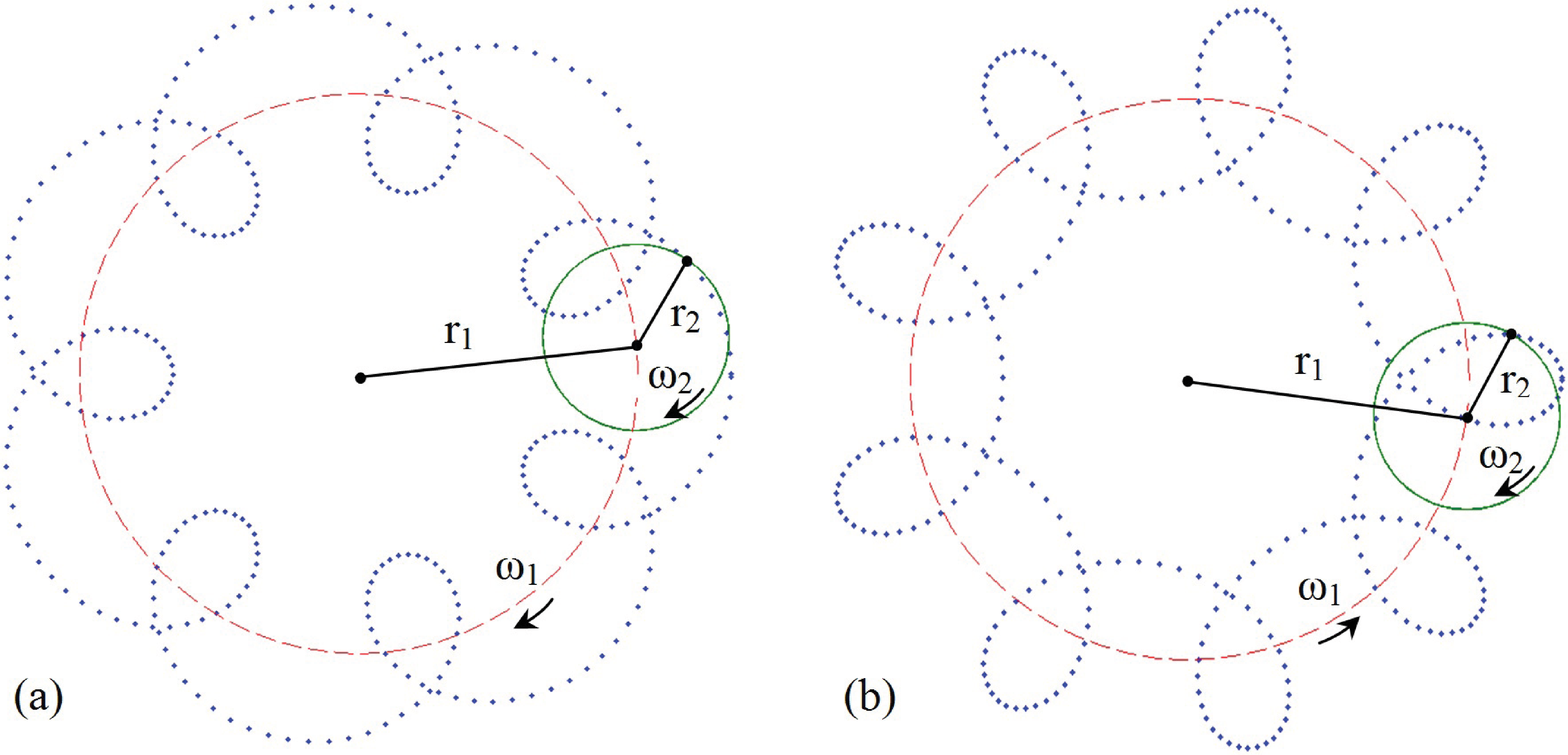}
  \caption{(Color online) Schematic diagram of cycloid motions. (a): epicycloid; (b): hypocycloid. Dotted line represents the trajectory of cycloid motion of grain. Solid line and dashed line indicate the primary circle and the secondary circle, respectively. $r_{2}$ and $r_{1}$ represent the radii of the primary and the secondary circles, respectively. $\omega_{2}$ and $\omega_{1}$ represent the angular velocities of the primary and the secondary circles, respectively. (a) and (b) occur when $\omega_{1}\omega_{2}$$>$$0$ and $\omega_{1}\omega_{2}$$<$$0$, respectively.}
  \end{center}
\end{figure}

\indent Circular motion is observed both inside and outside the glass ring as shown in Fig. 2(e). The angular speed of circular motion of grain changes within the range of 0.05-1.23 rad/s. Stationary grain is observed both inside and outside the glass ring as shown in Fig. 2(f). In this case the resultant force acting on the grain should be zero. In the absent of action with other grain, the grain is considerably stable at higher gas pressure. However, Brownian motion of grain is observed clearly at lower gas pressure. The grain moves stochastically near its equilibrium position. This has also been reported by other groups \cite{Schwabe2}.

\indent Which trajectory a grain will follow is determined by the initial conditions (such as the initial position) of the dropped grain and the discharge parameters. This means that for several grains dropped simultaneously (or one grain dropped repeatedly), they (it) will exhibit different trajectories because their (its) initial conditions are not exactly the same. Therefore, the dependencies of the parameters $r_{1,2}$ and $\omega_{1,2}$ mentioned above on the gas pressure and the forward power of discharge could not be illustrated evidently. In our experiments, the radii of the primary and the secondary circles change within the range of $r_{2}$$=$$0.08-0.57$ mm and $r_{1}$$=$$13-25$ mm, respectively. The corresponding values of angular velocities change within the range of $\omega_{2}$$=$$24-88$ rad/s and $\omega_{1}$$=$$0.03-0.63$ rad/s, respectively.

\indent The (cuspate) cycloid motions of grains follow the equations in polar coordinate:

\begin{eqnarray}
  r^{2} &=& r_{1}^{2} + r_{2}^{2} +2r_{1}r_{2}cos(\omega_{2}-\omega_{1})t, \\
  \theta &=& \omega_{1}t \pm arccos(\frac{r_{1}+r_{2}cos(\omega_{2}-\omega_{1})t}{r}),
\end{eqnarray}
where, "+" and "-" correspond the cases of $sin(\omega_{2}-\omega_{1})t>0$ and $sin(\omega_{2}-\omega_{1})t\leq0$, respectively. $r$ and $\theta$ represent the radial and azimuth positions of grain in polar coordinate, respectively. $\omega_{i}$ and $r_{i}$ ($i$$=$$1,2$) are defined as above. The observed motions of grains as shown in Fig. 2 can be reproduced well in term of the Eqs. (1)-(4).

\subsection{Force analysis }

\indent The force analysis for a nonspherical grain, like the pine pollen we used, becomes very complex. These forces mainly include the gravitation force, electrostatic force, ion drag force, neutral friction force, and Magnus force. The Magnus force, which originates from the spin of a grain, is considered here for the first time. The interaction force between grains is not considered in the present case because only one or few grains separated well are dropped in each experiment.

\indent The gravitation force $F_{g}$$=$$mg$ acts downward vertically and will not be considered here because our emphases is on the horizontal motion of the dust grain.

\indent The electrostatic force originates from the action of sheath field on the negatively-charged grain. In the present of glass ring the sheath profile follows approximately the wall of the glass ring and the electrode. When the diameter of the glass ring is small, the horizontal confinement to the grain could be described by a parabolic potential \cite{Fortov2}. The grains are confined in the center of the glass ring. Here, the glass ring is relatively large, and the grain is restricted near the glass wall both inside and outside. The horizontal confinement inside the glass ring is represented approximately by a quartic equation:

\begin{eqnarray}
  U(r) &=& \alpha(ar^{4} - br^{2}),
\end{eqnarray}
 here, $r$ represents the radial distance to the center of the glass ring, $\alpha$ controls the well depth, and $a$, $b$ control the well position in the radial direction. For the sake of simplicity, the horizontal confinement outside the glass ring has a mirror symmetry about the glass wall as shown by the solid lines in Fig. 5 (a). The upward vertical component of the sheath field acts against the downward gravitation force. The horizontal component of the sheath field plays important role on the complex motion of grain in horizontal plane.

\indent Ion drag force acting on the grain by ion flow results mainly from Coulomb drag and collection drag, and has been explored extensively. Together with the neutral drag and the plasma charging flux, the ion drag could give rise to the spin of the immersed grain. The spin of a hollow transparent glass microsphere with surface defect was observed in a stratified glow discharge by Karasev $et$ $al$ \cite{Karasev2, Karasev3}. The values of angular velocity measured in their experiments are between 0 and 12000 rad/s. As a rule, for the nonspherical grain as we used here, the pine pollen would spin faster than the glass microsphere they used because the surface of pine pollen is much irregular than that of glass microsphere. The rapid spin of the pine pollen around its center of inertia gives rise to the Magnus force which plays important role on the complex motions of grains.

\begin{figure}[htbp]
  \begin{center}\includegraphics[width=5cm,height=5cm]{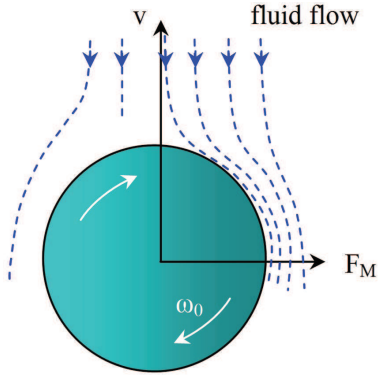}
  \caption{(Color online) Schematic diagram of the Magnus force $F_{M}$ acting on a spinning grain.}
  \end{center}
\end{figure}

\indent The Magnus force results from the spin of a grain as illustrated by Fig.4. In term of the Bernoulli principle, the pressure on the left of a spinning grain would be different from that on the right while the grain travelling forward. The pressure difference between the left and right sides gives rise to the Magnus force $F_{M}$:
\begin{eqnarray}
  \mathbf{F}_{M} &=& k \eta d^{3}\mathbf{\omega_{0}}\times\mathbf{v},
\end{eqnarray}
 here, $\mathbf{\omega_{0}}$ and $\mathbf{v}$ represent the vectors of angular velocity and travelling velocity of a spinning grain, respectively. $d$ is the diameter of grain. $\eta$ is the density of fluid. $k$ is a coefficient related with the shape of the grain. The direction of Magnus force acting on the grain is always perpendicular to the travelling direction of the grain, therefore the Magnus force provides the main centrifugal force for the cycloid motion of grain travelling along the primary circle. Here, we focus on the case that the vector of angular velocity of the spinning grain is along the vertical direction, i.e., the Magnus force acting on the grain is always along the horizontal plane. If we assume that the Magnus force equals approximately to the centrifugal force of the primary circle, then we have $k\omega_{0}$$\simeq$$\frac{m\omega_{2}}{\eta d^{3}}$$\simeq$$5$$\times$$10^{6}$ rad/s. Here, we suppose $k$$\simeq$$5\pi$ for the pine pollen, which is five times than that for the microsphere. Therefore, the average value of angular velocity of spin is about $\omega_{0}$$\simeq$3$\times$$10^{5}$ rad/s. This value is reasonably larger than that observed by Karasev $et$ $al$ because the pine pollen we used is more irregular than the hollow transparent glass microsphere they used. However, experimental detection of such a grain spinning faster is out of the reach of our camera at present. In order to confirm the distinctive features of the cycloid motions the pine pollen has exhibited as shown above, further experiments with glass and resin microspheres are performed at the same discharge conditions for comparison (not shown here). No cycloid motion is observed. Those regular microspheres show behaviours such as the random motion and the crystallization as other groups observed previously. Therefore, we can draw a conclusion that the spin of the pine pollen gives rise to the Magnus force which provides the centrifugal force for the primary circle.

\indent The neutral drag force is a resistance force caused by the collisions with the neutral gas molecules. It damps the motion of grain. Therefore a driving force is necessary for the grain to keep moving forward along the orbit. Although the origin of this unknown driving force is still unclear at present, we can give some suggestions. 1) It should be related with the irregularity of the grain, because our comparison experiments with glass and resin microspheres have shown that no cycloid motion and no circle motion are observed. Several groups also reported the rotation of nonspherical agglomerates of microspheres \cite{Paeva}. The rotation of microsphere is observed only in magnetized dust plasma, which is induced by the Lorentz force \cite{Karasev, Carstensen, Hou, Konopka}. 2) This driving force is nondirectional with regard to the glass ring because both clockwise and counterclockwise rotations are observed. 3) It should be the same order of magnitude of the neutral drag force because the grain could continue to move uniformly along the glass ring. In the following we will neglect the counteractive neutral drag force and driving force.

\begin{figure}[htbp]
  \begin{center}\includegraphics[width=8cm,height=8cm]{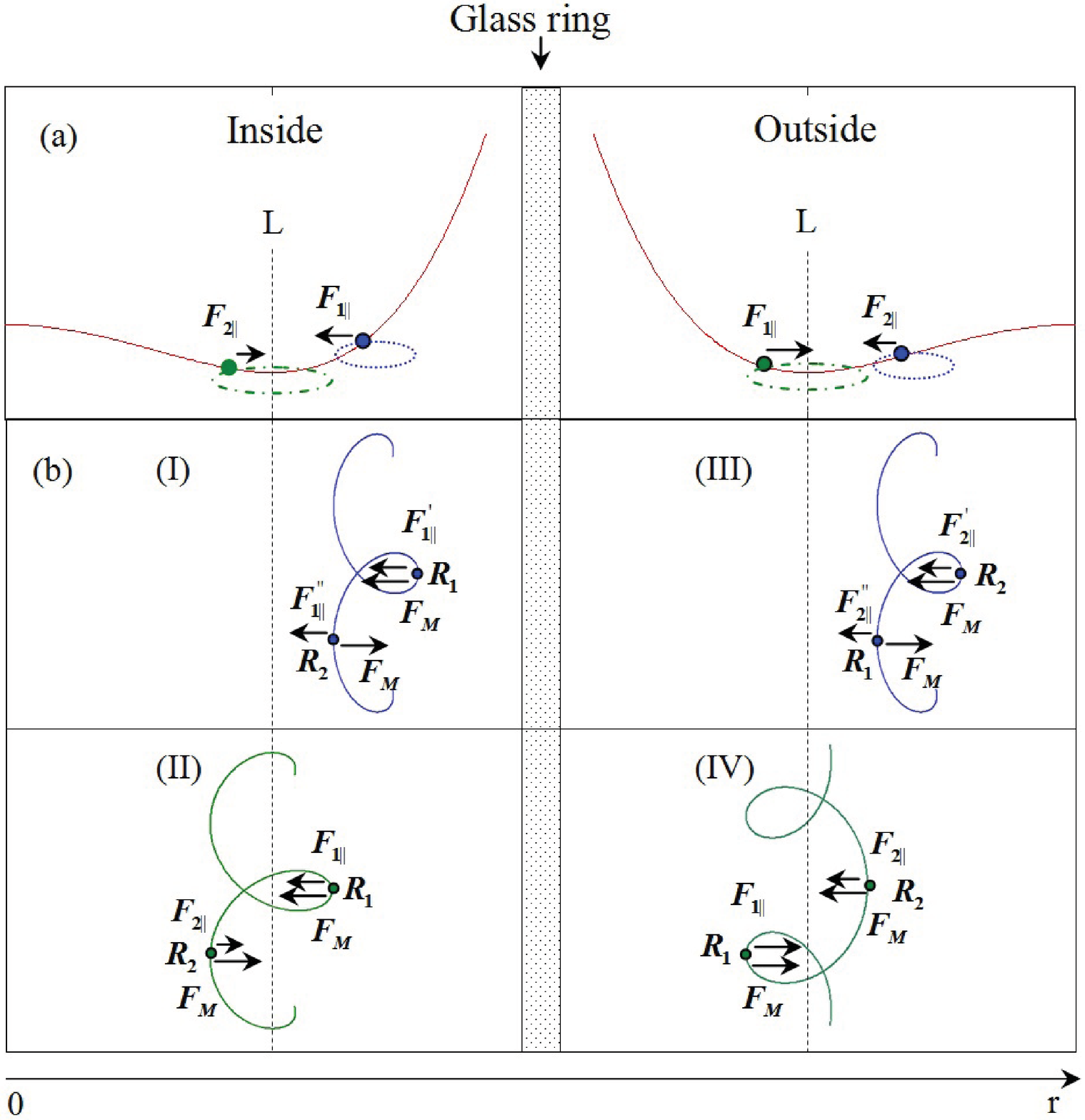}
  \caption{(Color online) Schematic diagram of the grain trajectories. Solid lines in (a) illustrate the approximate profile of sheath in vertical section. Equilibrium line $L$ (dashed line) indicates the position of the minima value of confinement potential. $F_{1\parallel}$ is the horizontal component of the resultant force a grain suffered when it is between the equilibrium line $L$ and the glass wall. Outside this region, it is expressed as $F_{2\parallel}$. (b) shows four cases of complex trajectories in horizontal plane. (I)-(III) show the hypocycloid motions, and (IV) shows the epicycloid motion. $F_{M}$ represents the Magnus force. $R_{1}$ ($R_{2}$) is the cyclotron radius when the grain is near (far away from) the glass wall.}
   \end{center}
\end{figure}

\indent In term of the above force analysis we explain next the complex motions of grains. For these regular trajectories observed in 2D horizontal plane we make the following assumptions for simplicity: 1) the vector of angular velocity of the spinning grain points downward, i.e., the grain spins clockwise; 2) the grain rotates clockwise around the primary circle, i.e., the value $\omega_{2}$$<$$0$; 3) the values of angular velocity $\omega_{0}$ and travelling velocity $v$ remain constant, i.e., the magnitude of Magnus force $F_{M}$ does not change. 4) the resultant force of the horizontal components of electrostatic force and ion drag force in radial direction is indicated by $F_{1\parallel}$ when the grain is located between the equilibrium line $L$ and the glass wall as shown in Fig. 5(a). Outside this region, it is expressed as $F_{2\parallel}$. In general, $F_{1\parallel}$$>$$F_{2\parallel}$ due to the steep sheath near the glass wall. Our observations of the (cuspate) cycloid motions of grains can be classified roughly into four groups (I)-(IV) as shown by the diagrams in Fig. 5 (b).

\indent For the case (I), the grain meanders between the equilibrium line $L$ and the glass wall. When the grain approaches the wall both the Magnus force $F_{M}$ and $F'_{1\parallel}$ act inwards. Far away from the glass wall, on the contrary, the two forces act in the opposite direction as shown in Fig. 5(b). The centrifugal forces in the two cases are:

\begin{eqnarray}
  F_{c1} &=& F_{M} + F'_{1\parallel} = \frac{mv^{2}}{R_{1}},\\
  F_{c2} &=& F_{M} - F''_{1\parallel} = \frac{mv^{2}}{R_{2}},
\end{eqnarray}
where, $R_{1}$ and $R_{2}$ represent the cyclotron radii near and far away from the glass wall, respectively. Therefore, the cyclotron radii $R_{1}$$<$$R_{2}$. The cyclotron radius changes periodically as the grain travelling, which results in the hypocycloid motion. If the rotation of grain is more closer to the glass wall, the forces $F'_{1\parallel}$ and $F''_{1\parallel}$ become stronger and the cyclotron radius $R_{1}$ ($R_{2}$) becomes more smaller (larger). This could lead to the formation of cuspate hypocycloid motion of grain as illustrated by Fig. 2(d). The above analysis is applicable to the case (III), in which the grain also follows (cuspate) hypocycloid trajectory.

\indent For the case (II), the grain meanders near the equilibrium line $L$. When the grain approaches the glass wall both the Magnus force $F_{M}$ and $F_{1\parallel}$ act inwards. When it is far away the glass wall, the two forces act outwards as shown in Fig. 5(b). The centrifugal forces in the two cases read:

\begin{eqnarray}
  F_{c1} &=& F_{M} + F_{1\parallel} = \frac{mv^{2}}{R_{1}},\\
  F_{c2} &=& F_{M} + F_{2\parallel} = \frac{mv^{2}}{R_{2}}.
\end{eqnarray}
Here, $F_{1\parallel}$ is slightly larger than $F_{2\parallel}$ due to the steep sheath near the glass wall, therefore the cyclotron radii $R_{1}$$<$$R_{2}$. Similarly, the periodical change of cyclotron radius gives rise to the hypocycloid trajectory as illustrated by Fig. 2(c).

\indent The above force analysis is also applicable to the case (IV). However, because the small cyclotron radius $R_{1}$ is located on the left of the equilibrium line $L$ in case (IV), the periodical change of cyclotron radius results in the epicycloid trajectory as illustrated by Fig. 2(a). Particularly, if the grain is more closer to the glass wall, $F_{1\parallel}$ becomes stronger, the petals would turn into cusps, which results in the formation of cuspate epicycloid trajectory as illustrated by Fig. 2(b).

\indent For stationary grain as shown in Fig. 2(f), the resultant force acting on it in horizontal plane should be zero. The grain should stay at the bottom of the sheath. Therefore, the position of the stationary grain can be used to indicate as the equilibrium line $L$. For the grain following circle trajectory as shown in Fig. 2(e), it locates on the outside of the equilibrium line $L$. The resultant force acting on it points to the center of the glass ring and plays the role of centripetal force which leads to the circle motion. In both cases, the grain could spin slightly, and the Magnus force is too small to give rise to the cycloid motion of grain.

\subsection{NUMERICAL SIMULATIONS}

\indent We consider the motion of a charged grain with mass $m$. The grain is confined in the potential $U$ as described by Eq. 5. In 2D Cartesian coordinate system the dynamic motion of the grain can be described by
\begin{eqnarray}
  m\frac{dV_{x}}{dt} &=& MV_{y} + F_{r\|}\frac{x}{r},\\
  m\frac{dV_{y}}{dt} &=& -MV_{x} + F_{r\|}\frac{y}{r},
\end{eqnarray}
where $r$$=$$\sqrt{x^{2}+y^{2}}$. $F_{r\|}$$=$$-\beta(2ar^{3}-br)$ represents the resultant force of the horizontal components of electrostatic force and ion drag force, in which $\beta$ defines the strength of the radial confinement. $M$$=$$k\eta d^{3}\omega_{0}$ is the coefficient in Magnus force. Here, we neglect the counteractive neutral drag force and driving force in both $x$ and $y$ directions. The average diameter of the grain is about $d$ $\simeq$ 4 $\times$ $10^{-5}$ m, and the mass of the grain is estimated to be of $m$ $\simeq$ 6.7 $\times$ $10^{-12}$ kg. In term of the position of the potential well which can be indicated experimentally by the stationary grain, the magnitudes of $a$ and $b$ are determined to be 85 and 0.03, respectively.

\begin{figure}[htbp]
  \begin{center}\includegraphics[width=8cm,height=3cm]{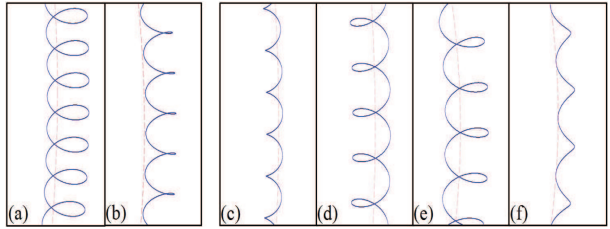}
  \caption{(Color online) Numerical results of the (cuspate) cycloid motions. (a)-(b): inside the glass ring; (c)-(f): outside the glass ring. Dashed line indicates the position of the minima value of confinement potential, i.e., the equilibrium line $L$ in Fig.5. The initial positions apart from the center of the glass ring in (a)-(f) are set to $x_{0}$$=$$13.24$, $13.28$, $24.47$, $24.34$, $25.5$, and $24.7$ mm, respectively. The corresponding initial speeds in (a)-(f) are  $V_{y}$$=$$0.004$, $0.004$, $-0.0008$, $0.001$, $-0.002$, $0.01$ m/s, respectively. Other initial parameters are the same in (a)-(f): $y_{0}$$=$$0.0$ mm, $V_{x}$$=$$0.0$ m/s, $\omega_{0}$$=$$2$$\times$$10^{5}$ rad/s, $k$$=$$15$, $\beta$$=$1$\times$$10^{7}$.}
  \end{center}
\end{figure}

\begin{figure}[htbp]
  \begin{center}\includegraphics[width=7cm,height=5cm]{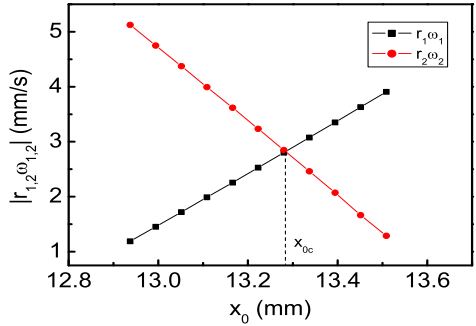}
  \caption{(Color online) Dependance of the linear velocities of the primary and the secondary circles on the initial position. Below (above) the critical initial position $x_{0c}$, hypocycloid (epicycloid) motion of grain occurs. Other initial parameters are the same as those in Fig. 6(a).}
  \end{center}
\end{figure}

\indent Figure 6 shows the numerical results based on Eqs. (11)-(12) with different initial conditions. It shows that when the grain is located near the equilibrium line one obtains the cycloid trajectories. However, far from the equilibrium line, the petals could disappear, which results in the formation of cuspate cycloid trajectories. These trajectories are in accordance with the experimental observations as shown in Fig. 2 and the force analysis above. The dependance of the magnitudes of linear speeds $|r_{i}$$\omega_{i}|$ (here, $i$$=$$1,2$) of the primary and secondary circles on the initial position $x_{0}$ is shown in Fig. 7. Below the critical position $x_{0c}$, $|r_{1}$$\omega_{1}|$$<$$|r_{2}$$\omega_{2}|$, one obtains hypocycloid motion. Above the critical position $x_{0c}$, $|r_{1}$$\omega_{1}|$$\geq$$|r_{2}$$\omega_{2}|$, one observes cuspate hypocycloid motion. These relations are in accordance with Eqs. (1)-(2).

\section{CONCLUSIONS AND REMARKS}
\indent In this work, we have studied the cycloid motion of an irregular grain (pine pollen) in unmagnetized dusty plasma in 2D horizontal plane. Hypocycloid motions are observed both inside and outside the glass ring. Epicycloid motion is observed only outside the glass ring. The stationary grain should stay at the bottom of the sheath and can be used to indicate the equilibrium line $L$. Circle motions are observed on the outside of the equilibrium line. The rapid spin of grain gives rise to Magnus force which plays the role of centrifugal force of the primary circle. The magnitude of angular velocity of spin is estimated to be of the order $\omega_{0}$$\sim$$10^{5}$ rad/s. Based on the force analysis and numerical simulations in 2D horizontal plane, the mechanisms of these cycloid motions are discussed. The observed cycloid motion of grain originates from the periodical change of cyclotron radius near and far from the glass wall.

\indent As is well known, the Magnus and Lorentz forces are characteristic of their transverse action with respect to the travelling direction of an object. No magnetic field is applied in our experiments. Therefore, only the Magnus force appears and provides the main centrifugal force for the rotation of grain along the primary circle. Recently, some studies focused on the Magnus force effect in optical manipulation \cite{Cipparrone}, in a Bose-Einstein Condensate \cite{Turner}, and in lattice models of superfluids \cite{Sonin}. They have revealed that the Magnus force is larger than that predicted by theoretical models and allows one to perform particle trapping. Our results have also revealed that the Magnus force an irregular grain suffered in plasma environment can not be negligible and would play important role on the complex motions of the grains. It is also suggested that these complex motions of irregular grain we observed could be found in fusion reactors because the grown or sputtered grain could have considerable angular speed of spin.

\indent In addition, the vector of angular velocity of a spinning grain is not limited to the downward direction as we discussed above. Due to the stochastic initial conditions of a dropped grain, this vector could point at other directions in three dimensional space. This means that the Magnus force acting on a spinning grain could point out of the horizontal plane. Therefore, the Magnus force together with the other forces will give rise to more complex motions of a spinning grain in three dimensional space. This will be the subject of a future study.

\section{ACKNOWLEDGMENTS}
\indent  This work is supported by the National Natural Science Foundation of China (Grant No. 11205044), the Natural Science Foundation of Hebei Province, China (Grants Nos. A2011201006, A2012201015), the Research Foundation of Education Bureau of Hebei Province, China (Grant No. Y2012009), and the Science Foundation of Hebei University.

\end{document}